\documentclass[12pt]{article}
\usepackage{graphicx}
\def \b{{\cal B}}
\def \bea{\begin{eqnarray}}
\def \beq{\begin{equation}}
\def \bo{B^0}
\def \bra#1{\langle #1 |}

\def \eea{\end{eqnarray}}
\def \eeq{\end{equation}}

\def \ket#1{| #1 \rangle}

\def \pr{\parallel}

\def \rpp{R_{\pi \pi}}

\def \s{\sqrt{2}}

\begin{document}
\begin{flushright}
EFI 03-25 \\
hep-ph/0305262 \\
May 2003 \\
\end{flushright}
\medskip

\centerline {\bf THE FACTORIZABLE AMPLITUDE IN $B^0 \to \pi^+ \pi^-$
\footnote{To be submitted to Physical Review D.}}
\bigskip

\centerline {Zumin Luo~\footnote{zuminluo@midway.uchicago.edu} and
Jonathan L. Rosner~\footnote{rosner@hep.uchicago.edu}}
\centerline {\it Enrico Fermi Institute and Department of Physics}
\centerline{\it University of Chicago, 5640 S. Ellis Avenue, Chicago, IL 60637}
\bigskip

\begin{quote}

Using the measured spectrum shape for $B \to \pi \ell \nu$, the
rate for $B^+ \to \pi^+ \pi^0$, information on the
Cabibbo-Kobayashi-Maskawa (CKM) matrix element $|V_{ub}|$, and
theoretical inputs from factorization and lattice gauge theory, we
obtain an improved estimate of the ``tree'' contribution to $B^0
\to \pi^+ \pi^-$.  We find the branching ratio $\b(B^0 \to \pi^+
\pi^-)|_{\rm tree} = (5.25^{+1.67}_{-0.50}) \times 10^{-6}$, to be
compared with the experimental value $\b(\bo \to \pi^+ \pi^-) =
(4.55 \pm 0.44) \times 10^{-6}$.  The fit implies $|V_{ub}| = (3.62
\pm 0.34) \times 10^{-3}$.  Implications for tree-penguin interference in $\bo
\to \pi^+ \pi^-$ and for other charmless $B$ decays are discussed.
\end{quote}
\bigskip

\noindent
PACS Categories:  13.25.Hw, 14.40.Nd, 14.65.Fy, 11.30.Er
\bigskip

\centerline{\bf I.  INTRODUCTION}
\bigskip

The semileptonic process $B \to \pi \ell \nu$ involves a form
factor $F_+(q^2)$ related for $q^2 = m_\pi^2$ to the factorized
color-favored ``tree'' contribution in $B^0 \to \pi^+ \pi^-$
\cite{Vol,BJ,BS}. In previous work \cite{bpilnu} we obtained an
estimate of this contribution implying a branching ratio $\b(\bo
\to \pi^+ \pi^-)|_{\rm tree} = (7.3 \pm 3.2) \times 10^{-6}$.  A
measurement of the spectrum $d \Gamma(B \to \pi \ell \nu)/dq^2$
has now been presented by the CLEO Collaboration \cite{CLEOsl03}
working at the Cornell Electron Storage Ring.  Further results are
expected from the BaBar and Belle Collaborations at asymmetric $e^+ e^-$
colliders.  Using the CLEO measurement and other inputs, we find in the present
paper an improved estimate implying $\b(\bo \to \pi^+ \pi^-)|_{\rm
tree} =(5.25^{+1.67}_{-0.50}) \times 10^{-6}$, to be compared with
the observed branching ratio $\b(\bo \to \pi^+ \pi^-) = (4.55 \pm
0.44) \times 10^{-6}$. This result has a number of implications
for tree-penguin interference in $\bo \to \pi^+ \pi^-$ and for
other charmless $B$ decays.

We review theoretical inputs, including constraints
from factorization and lattice gauge theory calculations, in Sec. II, while
data are discussed in Sec.\ III.  We perform a global fit to these inputs in
Sec.\ IV.  The consequences of this fit are discussed for $B^0 \to \pi^+ \pi^-$
and other charmless $B$ decays in Sec.\ V.  We conclude in Sec.\ VI.
\newpage

\centerline{\bf II.  THEORETICAL INPUTS}
\bigskip

The $B \to \pi$ matrix element is parametrized by two independent form factors:
\beq
\bra{\pi(p)} \bar{u}\gamma_{\mu}b \ket{B(p+q)} =
\left(2p+q-q\frac{m_B^2-m_{\pi}^2}{q^2}\right)_{\mu}F_+(q^2) +
q_{\mu}\frac{m_B^2-m_{\pi}^2}{q^2}F_0(q^2)~ ,
\eeq
For massless leptons (assumed here), only $F_+(q^2)$ contributes to the
differential decay rate
\beq \label{eqn:diff}
\frac{d\Gamma}{dq^2}(B^0 \to \pi^-\ell^+ \nu_{\ell}) =
\frac{G_F^2|V_{ub}|^2}{24\pi^3}|\vec{p}_{\pi}|^3|F_+(q^2)|^2~~ ,
\eeq
where $V_{ub}$ is the relevant CKM matrix element.  In the factorization
hypothesis, one replaces the lepton pair with a pion, giving what we term
the ``tree'' contribution $T$ (in the notation of \cite{GHLR}) to
the nonleptonic decay $B^0 \to \pi^+ \pi^-$. In the limit of small
$m_{\pi}$, the two processes are related by \cite{BJ}
\beq \label{eqn:pipi}
\Gamma_{\mathrm{tree}}(B^0 \to \pi^+
\pi^-)=6\pi^2f_{\pi}^2|V_{ud}|^2|a_1|^2\left.\frac{d\Gamma(B^0 \to \pi^-
\ell^+ \nu_{\ell})}{dq^2}\right|_{q^2=m_{\pi}^2} .
\eeq
where $|a_1|$ is a QCD correction which we shall take equal to 1.
The majority of QCD effects are expected to be associated with the form
factors $F_{+,0}(q^2)$ and thus are taken into account by the factorization
{\it ansatz}.

Other contributions to charmless strangeness-preserving $B$ decays which we
shall consider include color-suppressed tree ($C$) and penguin ($P$)
amplitudes.  The corresponding strangness-changing amplitudes are denoted
by primes.  We neglect smaller amplitudes which involve spectator quarks.
For these and other details, see, e.g., Ref.\ \cite{GHLR}.  The amplitude
for $B^0 \to \pi^+ \pi^-$ is then
\beq
A(B^0 \to \pi^+ \pi^-) = - (T + P) = - |T|e^{i \gamma} - |P|e^{-i \beta}
e^{i \delta}~~~,
\eeq
where we have introduced phases of CKM elements, assuming the phase of the
$\bar b \to \bar d$ penguin to be dominated by the top quark, and $\delta$
denotes a relative strong phase.  A question which has been of interest
for some time \cite{bpilnu,GRdest,Hdest} is whether the small branching ratio
for $B^0 \to \pi^+ \pi^-$ reflects the effect of destructive tree-penguin
interference.  If so, by combining this information with CP-violating
asymmetries in $B^0 \to \pi^+ \pi^-$, one can learn a good deal about both weak
(i.e., CKM) and strong phases \cite{GRpipi}.

We use notation in which the square of an amplitude directly gives a $B^0$
branching ratio in units of $10^{-6}$.  The observed branching ratio
$\b(B^0 \to \pi^+ \pi^-) = (4.55 \pm 0.44) \times 10^{-6}$ (see Sec.\ III)
then corresponds to $|T+P| = 2.13 \pm 0.10$ in our units.  In previous work
\cite{bpilnu} we found $|T| = 2.7 \pm 0.6$, too large an error to display
any possible tree-penguin interference.

Another amplitude which will be of use to us is $A(B^+ \to \pi^+ \pi^0) = -
(T+C)/\s$.  The color-suppressed amplitude $C$ is expected to have small phase
and magnitude relative to $T$ \cite{BBNS}.  We shall use only the conservative
range \cite{MN} $0.08 < |C/T| < 0.37$.  This will provide a useful bound on $T$
based on $\b(B^+ \to \pi^+ \pi^0)$.

Lattice gauge theories predict not only the shape, but also the
normalization, of the $B \to \pi$ form factors at large $q^2$ or
small pion recoil momentum in the $B$ rest frame \cite{UKQCD, APE,
FNAL, JLQCD}. These predictions turn out to be very helpful in
constraining parameters on the basis of the $q^2$ spectrum in $B
\to \pi l \nu$.  However, they do not address the key question of
the form factor behavior at small $q^2$ or large pion recoil.

The CKM parameter $V_{ub}$ is another key input whose determination is for the
moment still subject to theoretical uncertainties.  Good understanding of
the $B \to \pi$ form factor would reduce these uncertainties.  Independently
of $B \to \pi l \nu$ (or the more complex process $B \to \rho l \nu$), however,
various inclusive methods have been employed to extract $V_{ub}$ from
semileptonic $b \to u$ decays, including the study of leptons with energy
exceeding the endpoint for $b \to c l \nu$, the rejection of events with
recoil mass above charm threshold, and the use of the photon energy
distribution in $b \to s \gamma$ to measure the ``Fermi distribution'' of $b$
quarks inside a $B$ meson.  These are summarized in a subsection of the
Review of Particle Physics \cite{BG}.

The form factor $F_+(q^2)$ is expected to have a $B^*$ pole, as
well as possible higher-lying poles in $q^2$.  In Ref.\
\cite{bpilnu} we approximated it with a dipole form proposed by
Becirevic and Kaidalov \cite{BK} on the basis of lattice gauge
theory calculations:
\beq F_+(q^2) =
\frac{c_B(1-\alpha_B)}{(1-q^2/m_{B^*}^2)(1-\alpha_B
q^2/m_{B^*}^2)}~~~. \eeq
A value of $\alpha_B$ between 0 and 1 would correspond to a pole
lying above $m_{B^*}^2$.
However, we were unable to achieve a good fit to the CLEO $B \to
\pi l \nu$ spectrum with this form. (The $\chi^2$ of the fit is
more than 3 for one degree of freedom.) A generalization of the
above form factor is to multiply it by $(1+a q^2/m_{B^*}^2)$,
where $a$ is an additional parameter. The resulting form factor is
equivalent to an explicit dipole:
\beq F_+(q^2) = \frac{R_1}{1-q^2/m_{B^*}^2} +
\frac{R_2}{1-\alpha_B q^2/m_{B^*}^2}~~~. \eeq
However, we were unable to achieve any fit for any physical
$\alpha_B$ to the numerical inputs in Sec.\ III
which represented any improvement over the single pole with this
form. We thus choose instead the two-parameter form
\beq F_+(q^2) = \frac{F(0)}{1-q^2/m_{B^*}^2}(1 + a
q^2/m_{B^*}^2)~~~. \eeq

\bigskip

\centerline{\bf III.  NUMERICAL INPUTS}
\bigskip

We summarize some information on $B \to \pi \pi$ branching ratios
\cite{Babrs,Bebrs,CLbrs} in Table I.  The central value of the
$\pi^+ \pi^0$ branching ratio exceeds that of $\pi^+ \pi^-$
despite the fact that the coefficient of its dominant $T$ term is
divided by $\s$ (see Sec.\ II).  To extract an amplitude for
comparison with $B^0$ decays we must first divide all $B^+$
branching ratios by the $B^+/B^0$ lifetime ratio \cite{LEPBOSC}
$\tau_+/\tau_0 = 1.073 \pm 0.014$ .

\begin{table}
\caption{Branching ratios for some charmless two-body $B$ decays, in units of
$10^{-6}$.
\label{tab:brs}}
\begin{center}
\begin{tabular}{|c|c|c|c|c|} \hline \hline
Mode & BaBar \cite{Babrs} & Belle \cite{Bebrs} & CLEO \cite{CLbrs} & Average \\
\hline
$\pi^+ \pi^-$ & $4.7 \pm 0.6 \pm 0.2$ & $4.4 \pm 0.6 \pm 0.3$ &
 $4.5^{+1.4+0.5}_{-1.2-0.4}$ & $4.55 \pm 0.44$ \\
$\pi^+ \pi^0$ & $5.5 \pm 1.0 \pm 0.6$ & $5.3 \pm 1.3 \pm 0.5$ &
 $4.6^{+1.8+0.6}_{-1.6-0.7}$ & $5.27 \pm 0.79$ \\ \hline \hline
\end{tabular}
\end{center}
\end{table}

Our estimate of $|T|$ based on $B^+ \to \pi^+ \pi^0$ then proceeds as follows.
After correcting for the lifetime ratio, we find $|T+C| = 3.13 \pm 0.24$.
With \cite{MN} $|T+C| = |T|(1.23 \pm 0.15)$ we then obtain $|T| = 2.55 \pm
0.37$.  This is consistent with our previous determination \cite{bpilnu} but
with smaller errors.  (The estimate $|T| = 3.0 \pm 0.3$ of Ref.\ \cite{xiao}
uses too restrictive a value of $|C/T|$ in our opinion.)  We seek further
information from the $B \to \pi l \nu$ spectrum shape and other sources.  This
value would imply $\b(\bo \to \pi^+ \pi^-)|_{\rm tree} = (6.5 \pm 1.9) \times
10^{-6}$, only $1 \sigma$ above the experimental branching ratio.

The CLEO Collaboration \cite{CLEOsl03} measured $d \b(B \to \pi l \nu)
/dq^2$ in three $q^2$ bins, each about 8 GeV$/c^2$ wide.  The results are not
very sensitive to the choice of form factor and we quote them for the form
factor \cite{Ball01} which appears to fit the data best:
\beq \int dq^2 \frac{d \b}{d q^2}(B^0 \to \pi^- l^+ \nu_l) =
\left\{
\begin{array}{l l} (0.431 \pm 0.106) \times 10^{-4} & (0 \le q^2 \le
8~{\rm GeV}^2)
\\ (0.651 \pm 0.105) \times 10^{-4} & (8 \le q^2 \le 16~{\rm GeV}^2)
\\ (0.245 \pm 0.094) \times 10^{-4} & (16~{\rm GeV}^2 \le q^2)
\end{array} \right.
\eeq
Only statistical errors (dominant) are shown.  These sum to a
branching ratio of $\b(B^0 \to \pi^- l^+ \nu_l) = (1.33 \pm 0.18
\pm 0.11 \pm 0.01 \pm 0.07) \times 10^{-4}$, where the errors are
statistical, experimental systematic, pion form factor
uncertainty, and $\rho$ form factor uncertainty.

Lattice calculations of the form factor $F_+(q^2)$ have been
presented in the past few years by the UKQCD \cite{UKQCD}, APE
\cite{APE}, Fermilab \cite{FNAL} and JLQCD \cite{JLQCD}
Collaborations. Numerical values of $F_+(q^2)$ are computed in the
range $13.6\ {\rm GeV}^2 \le q^2 \le 23.4\ {\rm GeV}^2$. Although
small variations are present among the four different
calculations, all results are consistent with each other within
errors. We will include them all in our fits.

For $|V_{ub}|$ we combine determinations presented in Ref.\
\cite{BG} in the following manner.  All numbers will be quoted in
units of $10^{-3}$.  The inclusive LEP average is $4.09 \pm 0.37
\pm 0.44 \pm 0.34$ while the inclusive CLEO value is $4.12 \pm
0.34 \pm 0.44 \pm 0.33$, where the errors are statistical,
experimental systematic, $b \to c$ uncertainty, and $b \to u$
uncertainty.  In addition there are theoretical uncertainties
estimated to range up to 15\%.  Combining the two inclusive
numbers before folding in the theoretical uncertainties, and
treating the last two errors as common, we obtain $4.11 \pm 0.61$.
We shall use this in our fits. We do not include some preliminary results
presented by CLEO \cite{Bornheim:2002jk} and Belle \cite{Limosani:2003cc}.

An earlier CLEO exclusive determination of $|V_{ub}|$ utilizes
both $\pi l \nu$ and $\rho l \nu$ decays \cite{BG}.  Its result,
which we do not use in the present fit, amounts to an average of $3.25$
with experimental and theoretical errors comparable to those in the
inclusive determinations. Averaging it with the inclusive value noted
above, we should expect a global fit to give $10^3~|V_{ub}| \simeq 3.68
\pm 0.43$ with an additional 15\% theoretical error, or approximately a
20\% error overall.  We shall see that a modest improvement upon this
error is possible, while the central value does not change much.
\newpage

\centerline{\bf IV.  GLOBAL FIT}
\bigskip

We perform an overall three-parameter $\chi^2$ fit to the
above-mentioned $B \to \pi l \nu$ branching fractions in the three
$q^2$ bins, the averaged inclusive $|V_{ub}|$, and 26 lattice data
points on $F_+(q^2)$. We neglect the small correlations among the
three branching fractions. The quality of the fit is fairly good,
with $\chi^2 = 8.7$ for 27 degrees of freedom. The $\chi^2$'s
contributed by specific sources are summarized in Table
\ref{tab:chisq}. More than 50\% of the $\chi^2$ comes from the
Fermilab lattice points, which appear to be of a somewhat
different pattern from the other three lattice determinations.

\begin{table}
\caption{Sources of $\chi^2$ in global fit to $F_+(q^2)$.
\label{tab:chisq}}
\begin{center}
\begin{tabular}{c c c} \hline \hline
Source & $\chi^2$ & Reference \\ \hline
$B \to \pi l \nu$ spectrum & 2.42 & \cite{CLEOsl03} \\
Inclusive $|V_{ub}|$ values & 0.65 & \cite{BG} \\
UKQCD lattice data & 0.74 & \cite{UKQCD}\\
APE lattice data & 0.12 & \cite{APE}\\
Fermilab lattice data & 4.53 & \cite{FNAL} \\
JLQCD lattice data & 0.21 & \cite{JLQCD}\\ \hline \hline
\end{tabular}
\end{center}
\end{table}

\begin{figure}
\begin{center}
\includegraphics[height=7.5in]{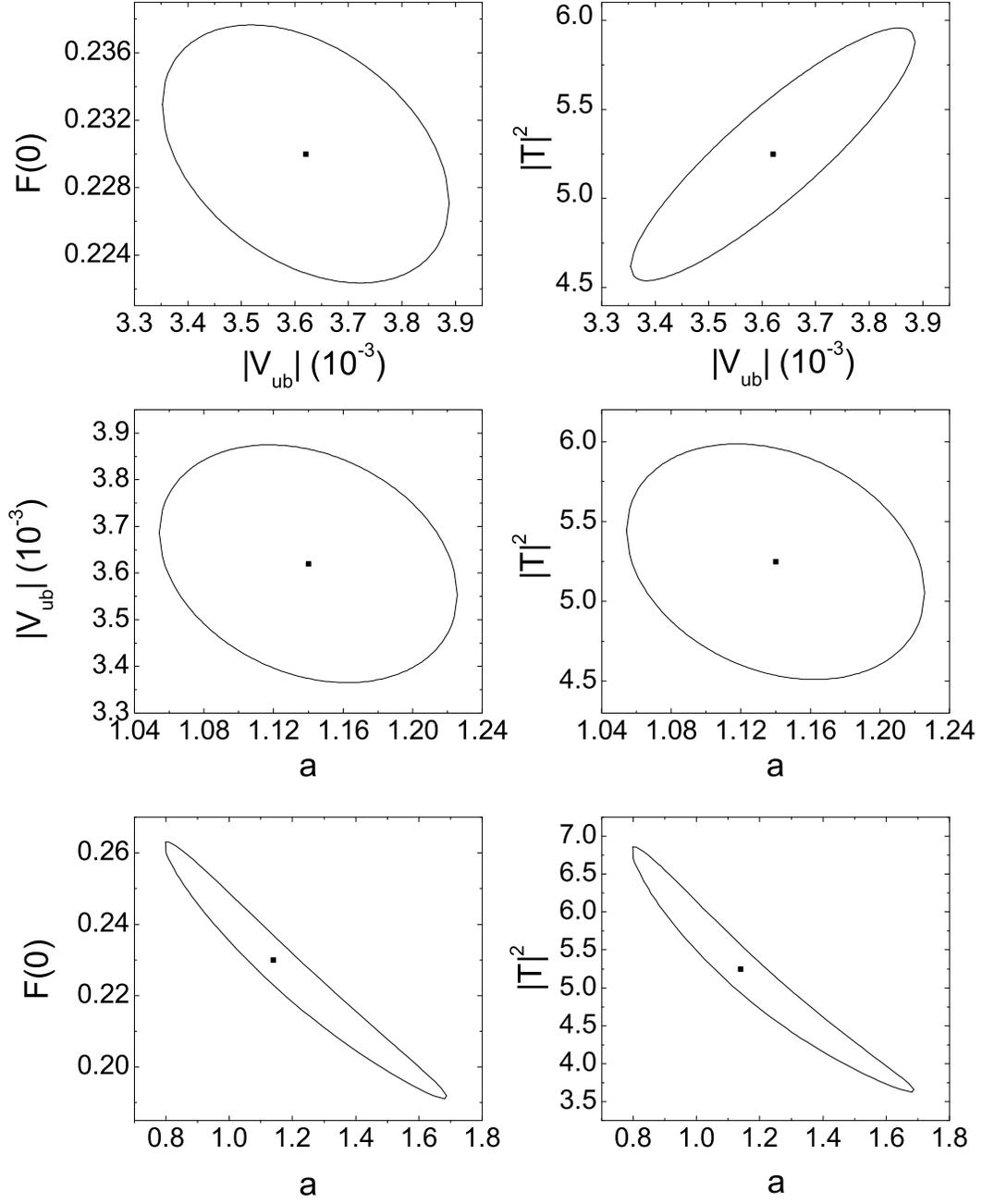} \\
\caption{Left column: projections of error ellipsoid for global fit on the
plane of two parameters for central values of the third. Right
column: ellipses involving the tree amplitude corresponding to the
variations shown in the left column. Note that $|T|^2 =
\b(B^0 \to \pi^+ \pi^-)|_{\rm tree}$ in units of $10^{-6}$.
\label{fig:ellipses}}
\end{center}
\end{figure}

The results of the fit are
\bea
 a     & = & 1.14^{+0.72}_{-0.42}~~,\\
 F(0) & = & 0.23 \pm 0.04~~,\\
 |V_{ub}| & = & (3.62 \pm 0.34) \times 10^{-3}~~,\\
 \b(\bo \to \pi^+ \pi^-)|_{\rm tree} & = & (5.25 \pm 1.67) \times 10^{-6} ~~,\\
 |T| & = & 2.29 \pm 0.36~~.
\eea
A theoretical error of $\simeq 15\%$ must be added to $|V_{ub}|$.
The value of $|T|$ overlaps that ($|T| = 2.55 \pm 0.37$) obtained
in Sec.\ III from $B^+ \to \pi^+ \pi^0$, but the $\pi^+ \pi^0$
value indicates that $\b(\bo \to \pi^+ \pi^-)|_{\rm tree}$ is no
smaller than $4.75 \times 10^{-6}$.  Hence we shall truncate our
parameter space at this lower limit, and quote
\bea \b(\bo \to \pi^+ \pi^-)|_{\rm tree} & = &
(5.25^{+1.67}_{-0.50}) \times 10^{-6} ~~,\\ |T| & = &
2.29^{+0.36}_{-0.11}~~. \eea
The ranges of parameters contributing to the global fit are
illustrated in Fig.~\ref{fig:ellipses}, where we show the points
corresponding to minimum $\chi^2 = 8.7$ and the ellipses
corresponding to $\Delta \chi^2 = 1$. The various projections are
helpful in visualizing the full range of parameter variation.  In
particular, the value of $\b(\bo \to \pi^+ \pi^-)|_{\rm tree}$ can
vary substantially as a result of the uncertainty in $F(0)$, which
is still not well constrained by the data.  However, its $1
\sigma$ upper limit of $6.92 \times 10^{-6}$ is well below that
implied by the previous estimate of Ref.\ \cite{bpilnu}.

\bigskip

\begin{figure}
\begin{center}
\includegraphics[height=4.4in]{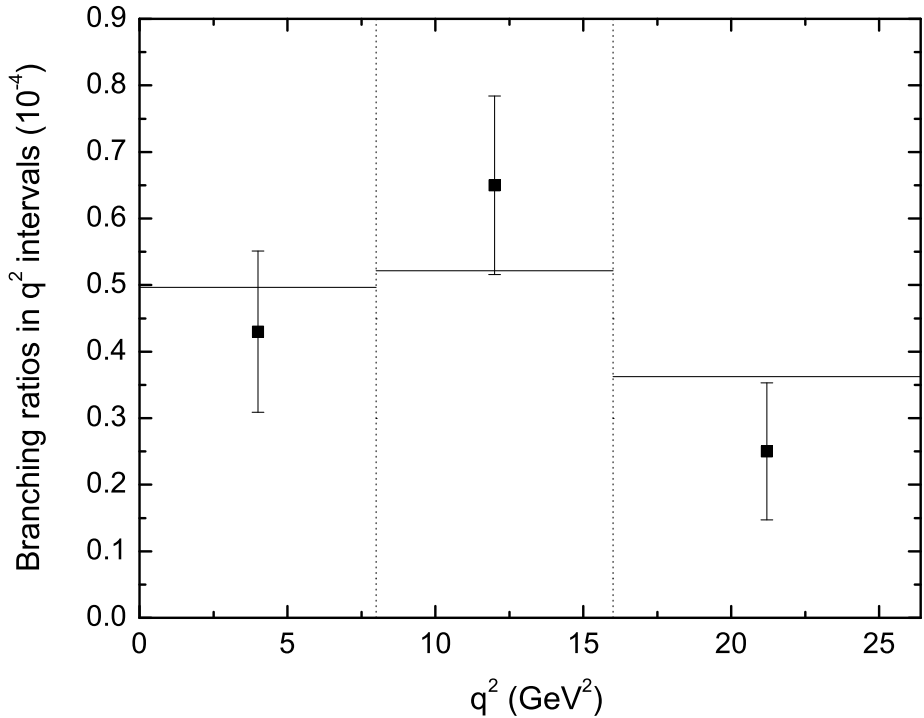} \\
\caption{Fit to $\int dq^2 \frac{d \b}{d q^2}(B^0 \to \pi^- l^+
\nu_l)$ values obtained for three $q^2$ bins in Ref.\
\cite{CLEOsl03}. \label{fig:br}}
\end{center}
\end{figure}

\begin{figure}
\begin{center}
\includegraphics[height=4.4in]{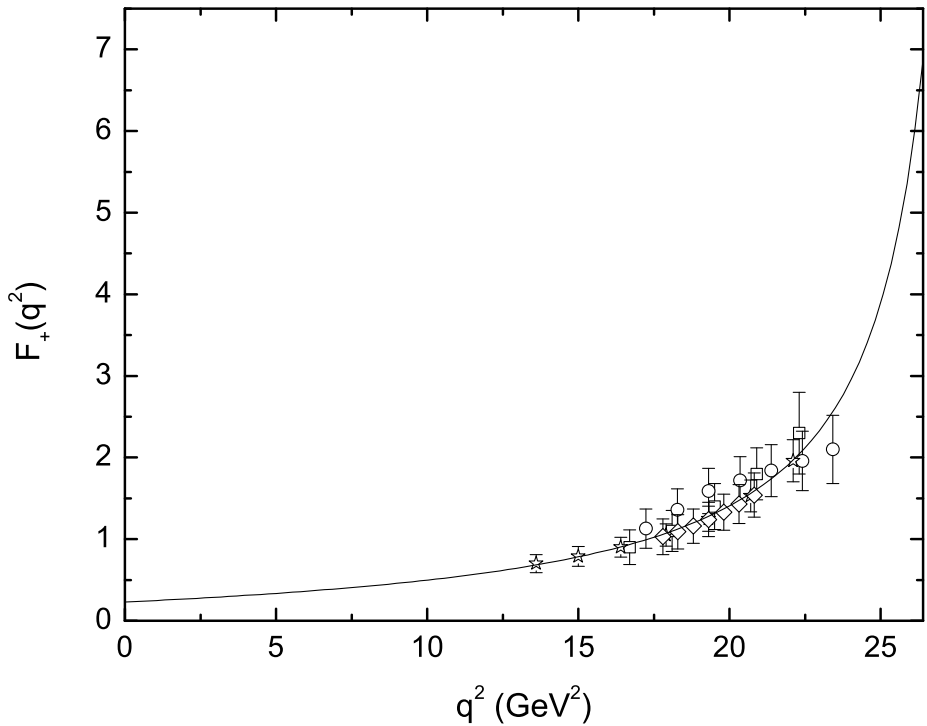} \\
\caption{Comparison of lattice data points with our best-fit form
factor $F_+(q^2)$. Lattice data are from UKQCD (squares), APE
(stars), Fermilab (circles) and JLQCD (diamonds). \label{fig:ff}}
\end{center}
\end{figure}

In Fig.~\ref{fig:br} we show our best fit to the CLEO data
\cite{CLEOsl03} for the $B^0 \to \pi^- l^+ \nu_l$ spectrum (in
three $q^2$ bins). The data favor a rather lower value of $F(0)$
than in our previous discussion \cite{bpilnu}, accounting for the
lower magnitude of the tree amplitude in the present treatment. In
Fig.~\ref{fig:ff} we show the comparison of the lattice data
points with our best-fit form factor $F_+(q^2)$.
As a consequence of the internal variations within the lattice
results, a $\chi^2$ of about 5.5 (contributed by the lattice data)
should be common for all fits; see Table II. Therefore, since the
$B \to \pi l \nu$ spectrum is the second largest $\chi^2$ source,
a significantly better overall fit can be achieved only if the
measured $B \to \pi l \nu$ branching ratios in the three $q^2$
bins are fitted better. This will require the addition of a fourth
parameter to affect the shape of $d \b(B \to \pi l \nu) /dq^2$ so
that it is suppressed at both low and high $q^2$ ends and enhanced
in the middle while relatively unchanged in the region $13.6\ {\rm
GeV}^2 \le q^2 \le 23.4\ {\rm GeV}^2$ where lattice data exist;
see Fig.~\ref{fig:br}. Consequently smaller tree amplitudes are
implied and we regard them as disfavored by the lower limit as
obtained earlier.

\bigskip

\centerline{\bf V.  HOW KNOWING THE TREE AMPLITUDE HELPS}
\bigskip

The ratio $\rpp$ of the observed $B^0 \to \pi^+ \pi^-$ branching ratio to its
value in the presence of the tree amplitude alone helps to establish the
relative magnitude and strong and weak phase of the penguin amplitude in this
process \cite{GRpipi}.  On the basis of the previous determination of the tree
amplitude \cite{bpilnu} and the present world average for $\b(B^0 \to \pi^+
\pi^-)$ we quoted \cite{mor} $\rpp = 0.62 \pm 0.28$, which indicated that
tree-penguin interference was not required but, if present in the rate, would
be destructive.  The new information on $|T|$ allows us to refine this estimate
to obtain $\rpp =0.87^{+0.11}_{-0.28}$, a value still consistent with both
possibilities.

The ratio $|P/T|$ of penguin to tree amplitudes quoted in Ref.\
\cite{mor} was $|P/T| = 0.28 \pm 0.06$.  This ratio is useful in
interpreting CP-violating asymmetries in the decay $B^0 \to \pi^+
\pi^-$ (see, e.g., \cite{GRpipi}). With the new world average
\cite{mor} $\b(B^+ \to K^0 \pi^+) = (19.6 \pm 1.4) \times 10^{-6}$
and the prescription \cite{GRpipi} $|P/P'| = (f_\pi/f_K)
\lambda/(1 - \lambda^2/2)$ we find for $f_\pi = 130.7$ MeV, $f_K =
159.8$ MeV, and \cite{Battaglia} $\lambda = 0.224$ the values
$|P'| = 4.28 \pm 0.16$, $|P| = 0.80 \pm 0.03$, and $|P/T| =
0.35^{+0.02}_{-0.06}$.  (Here the prime denotes a $|\Delta S| = 1$
amplitude.)  The ``penguin pollution'' thus is slightly greater
than estimated previously.  Corrections to the CKM phase $\alpha$
obtained from the asymmetry parameter $S_{\pi \pi}$ and the direct
asymmetry parameter $A_{\pi \pi}$ both can be slightly larger than
in Refs.\ \cite{GRpipi} and \cite{mor}.

The tree/penguin ratio in $B^0 \to K^+ \pi^-$ is also affected.  By a similar
analysis we found $r = |T'/P'| = 0.173 \pm 0.039$ in Ref.\ \cite{mor}; the new
value is $0.151^{+0.024}_{-0.009}$.  A bound on the CKM phase $\gamma$ quoted
in Ref.\ \cite{mor} relied on the lower limit of $r$, which is slightly raised,
so the bound is strengthened slightly.  Since it was only at the $1 \sigma$
level, we do not present it here.

A further implication of the improved upper bound on $T$ is a lower bound on
$C$.  Given the $1 \sigma$ bound $|T+C| \ge 2.89$ based on the $B^+ \to \pi^+
\pi^0$ branching ratio (see Sec.\ III) and the $1 \sigma$ upper bound $|T|
\le 2.65$ based on the present analysis, we conclude that if $C$ and $T$
have a small relative phase \cite{BBNS}, then Re($C/T) \stackrel{>}{\sim}
0.1$.
\bigskip

\centerline{\bf VI.  CONCLUSIONS}
\bigskip

The measurement of the $B \to \pi \ell \nu$ spectrum by the CLEO
Collaboration \cite{CLEOsl03} has provided valuable information
allowing us to improve the determination of the ``tree''
contribution to $B^0 \to \pi^+ \pi^-$.  Combining this information
with inclusive determinations of the CKM matrix element $|V_{ub}|$
and lattice gauge theory calculations of the $B \to \pi$ form
factor $F_+(q^2)$, we have found $\b(\bo \to \pi^+ \pi^-)|_{\rm
tree} = (5.25^{+1.67} _{-0.50}) \times 10^{-6}$, not significantly
greater than the experimental value $\b(\bo \to \pi^+ \pi^-) =
(4.55 \pm 0.44) \times 10^{-6}$.  The fit implies $|V_{ub}| = (3.62 \pm 0.34)
\times 10^{-3}$, with an additional theoretical error of 15\%.  The relative
strength of the penguin amplitude in this process, gauged using flavor SU(3)
from the rate for $B^+ \to K^0 \pi^+$, is slightly larger than
estimated previously, amounting to $(35^{+2}_{-6})\%$ in
amplitude. However, the need for strong destructive interference
between this amplitude and the tree contribution is somewhat
diminished in comparison with earlier estimates.
\bigskip

\centerline{\bf ACKNOWLEDGMENTS}
\bigskip

We thank Michael Gronau for helpful discussions.
This work was supported in part by the United
States Department of Energy through Grant No.\ DE FG02 90ER40560.

\def \ajp#1#2#3{Am.\ J. Phys.\ {\bf#1}, #2 (#3)}
\def \apny#1#2#3{Ann.\ Phys.\ (N.Y.) {\bf#1}, #2 (#3)}
\def \app#1#2#3{Acta Phys.\ Polonica {\bf#1}, #2 (#3)}
\def \arnps#1#2#3{Ann.\ Rev.\ Nucl.\ Part.\ Sci.\ {\bf#1}, #2 (#3)}
\def \art{and references therein}
\def \cmts#1#2#3{Comments on Nucl.\ Part.\ Phys.\ {\bf#1}, #2 (#3)}
\def \cn{Collaboration}
\def \cp89{{\it CP Violation,} edited by C. Jarlskog (World Scientific,
Singapore, 1989)}
\def \efi{Enrico Fermi Institute Report No.\ }
\def \epjc#1#2#3{Eur.\ Phys.\ J. C {\bf#1}, #2 (#3)}
\def \f79{{\it Proceedings of the 1979 International Symposium on Lepton and
Photon Interactions at High Energies,} Fermilab, August 23-29, 1979, ed. by
T. B. W. Kirk and H. D. I. Abarbanel (Fermi National Accelerator Laboratory,
Batavia, IL, 1979}
\def \hb87{{\it Proceeding of the 1987 International Symposium on Lepton and
Photon Interactions at High Energies,} Hamburg, 1987, ed. by W. Bartel
and R. R\"uckl (Nucl.\ Phys.\ B, Proc.\ Suppl., vol.\ 3) (North-Holland,
Amsterdam, 1988)}
\def \ib{{\it ibid.}~}
\def \ibj#1#2#3{~{\bf#1}, #2 (#3)}
\def \ichep72{{\it Proceedings of the XVI International Conference on High
Energy Physics}, Chicago and Batavia, Illinois, Sept. 6 -- 13, 1972,
edited by J. D. Jackson, A. Roberts, and R. Donaldson (Fermilab, Batavia,
IL, 1972)}
\def \ijmpa#1#2#3{Int.\ J.\ Mod.\ Phys.\ A {\bf#1}, #2 (#3)}
\def \ite{{\it et al.}}
\def \jhep#1#2#3{JHEP {\bf#1}, #2 (#3)}
\def \jpb#1#2#3{J.\ Phys.\ B {\bf#1}, #2 (#3)}
\def \lg{{\it Proceedings of the XIXth International Symposium on
Lepton and Photon Interactions,} Stanford, California, August 9--14 1999,
edited by J. Jaros and M. Peskin (World Scientific, Singapore, 2000)}
\def \lkl87{{\it Selected Topics in Electroweak Interactions} (Proceedings of
the Second Lake Louise Institute on New Frontiers in Particle Physics, 15 --
21 February, 1987), edited by J. M. Cameron \ite~(World Scientific, Singapore,
1987)}
\def \kdvs#1#2#3{{Kong.\ Danske Vid.\ Selsk., Matt-fys.\ Medd.} {\bf #1},
No.\ #2 (#3)}
\def \ky85{{\it Proceedings of the International Symposium on Lepton and
Photon Interactions at High Energy,} Kyoto, Aug.~19-24, 1985, edited by M.
Konuma and K. Takahashi (Kyoto Univ., Kyoto, 1985)}
\def \mpla#1#2#3{Mod.\ Phys.\ Lett.\ A {\bf#1}, #2 (#3)}
\def \nat#1#2#3{Nature {\bf#1}, #2 (#3)}
\def \nc#1#2#3{Nuovo Cim.\ {\bf#1}, #2 (#3)}
\def \nima#1#2#3{Nucl.\ Instr.\ Meth. A {\bf#1}, #2 (#3)}
\def \np#1#2#3{Nucl.\ Phys.\ {\bf#1}, #2 (#3)}
\def \npbps#1#2#3{Nucl.\ Phys.\ B Proc.\ Suppl.\ {\bf#1}, #2 (#3)}
\def \os{XXX International Conference on High Energy Physics, Osaka, Japan,
July 27 -- August 2, 2000}
\def \PDG{Particle Data Group, D. E. Groom \ite, \epjc{15}{1}{2000}}
\def \pisma#1#2#3#4{Pis'ma Zh.\ Eksp.\ Teor.\ Fiz.\ {\bf#1}, #2 (#3) [JETP
Lett.\ {\bf#1}, #4 (#3)]}
\def \pl#1#2#3{Phys.\ Lett.\ {\bf#1}, #2 (#3)}
\def \pla#1#2#3{Phys.\ Lett.\ A {\bf#1}, #2 (#3)}
\def \plb#1#2#3{Phys.\ Lett.\ B {\bf#1}, #2 (#3)}
\def \pr#1#2#3{Phys.\ Rev.\ {\bf#1}, #2 (#3)}
\def \prc#1#2#3{Phys.\ Rev.\ C {\bf#1}, #2 (#3)}
\def \prd#1#2#3{Phys.\ Rev.\ D {\bf#1}, #2 (#3)}
\def \prl#1#2#3{Phys.\ Rev.\ Lett.\ {\bf#1}, #2 (#3)}
\def \prp#1#2#3{Phys.\ Rep.\ {\bf#1}, #2 (#3)}
\def \ptp#1#2#3{Prog.\ Theor.\ Phys.\ {\bf#1}, #2 (#3)}
\def \rmp#1#2#3{Rev.\ Mod.\ Phys.\ {\bf#1}, #2 (#3)}
\def \rp#1{~~~~~\ldots\ldots{\rm rp~}{#1}~~~~~}
\def \si90{25th International Conference on High Energy Physics, Singapore,
Aug. 2-8, 1990}
\def \slc87{{\it Proceedings of the Salt Lake City Meeting} (Division of
Particles and Fields, American Physical Society, Salt Lake City, Utah, 1987),
ed. by C. DeTar and J. S. Ball (World Scientific, Singapore, 1987)}
\def \slac89{{\it Proceedings of the XIVth International Symposium on
Lepton and Photon Interactions,} Stanford, California, 1989, edited by M.
Riordan (World Scientific, Singapore, 1990)}
\def \smass82{{\it Proceedings of the 1982 DPF Summer Study on Elementary
Particle Physics and Future Facilities}, Snowmass, Colorado, edited by R.
Donaldson, R. Gustafson, and F. Paige (World Scientific, Singapore, 1982)}
\def \smass90{{\it Research Directions for the Decade} (Proceedings of the
1990 Summer Study on High Energy Physics, June 25--July 13, Snowmass, Colorado),
edited by E. L. Berger (World Scientific, Singapore, 1992)}
\def \tasi{{\it Testing the Standard Model} (Proceedings of the 1990
Theoretical Advanced Study Institute in Elementary Particle Physics, Boulder,
Colorado, 3--27 June, 1990), edited by M. Cveti\v{c} and P. Langacker
(World Scientific, Singapore, 1991)}
\def \yaf#1#2#3#4{Yad.\ Fiz.\ {\bf#1}, #2 (#3) [Sov.\ J.\ Nucl.\ Phys.\
{\bf #1}, #4 (#3)]}
\def \zhetf#1#2#3#4#5#6{Zh.\ Eksp.\ Teor.\ Fiz.\ {\bf #1}, #2 (#3) [Sov.\
Phys.\ - JETP {\bf #4}, #5 (#6)]}
\def \zpc#1#2#3{Zeit.\ Phys.\ C {\bf#1}, #2 (#3)}
\def \zpd#1#2#3{Zeit.\ Phys.\ D {\bf#1}, #2 (#3)}

\end{document}